# Human Error in IT Security

**Visahl Samson David Selvam**
BSc Cyber Security
Edith Cowan University – PSB Academy, Singapore
vdavidse@our.ecu.edu.au

## Abstract

This paper details on the analysis of human error, an IT security issue, and a major threat to the company. The human is one of the weakest links in the cybersecurity chain however it is a fundamental constituent of the embodiment. Verizon's 2019 data breach investigation has reported 34% of the data breaches are due to the threat of human error. The human errors could range from being deliberate to an unintentionally laid threat in the cyber space. Lack of awareness over the organization's information security policies, ignorance over the common information security practices or negligence by the employees are few of the reasons of data breaches caused by human error. However, countermeasures such as the use of McCumber cube methodology could mitigate the risk of human error caused to the organization. Furthermore, educating the employees through SETA program, simulated attacks, etc., can prevent significant security threats. This report provides a sophisticated elucidation on the various human errors and the necessary measures to mitigate and control the menacing IT security issue.

## Human error

Throughout this technology age, from industrial to business, cyber security plays a critical role everywhere. Maintaining confidentiality of information in an enterprise helps it to work and maintains consistent operations.

The human is the weakest link in the cybersecurity chain from the very beginning. Also, a survey conducted by a reputed organization (Swinhoe, 2019) in 2019 proves the fact remains unchanged even now. In the field of information security, human error can be divided into two groups, either deliberate or unintentional. An intentional one may occur due to another threat known as an insider threat, which has certain motivations behind it, while an unintentional one has no motivation or pre-planning, which may be due to a number of reasons, such as not knowing how a particular technology works, or lack of awareness.

According to research conducted by Kaspersky (Kaspersky, 2017), in 2017 nearly 49% pf malware/virus attacks are consisting of human error as contributing factors. In these attacks 53% are due to careless/uninformed employees, 36% of social engineering/phishing attacks, 38% is accidental hardware loss by the employee. Furthermore, as per Verizon's 2019 data breach investigation report (Verizon, 2019), 34% of the data breaches are due to the threat of human error.

**Human error causes and impacts**

The above figures demonstrate how much human error contributes to data breaches. And now the companies need to figure out how these issues are going to happen and how badly they will impact it. Several reasons can cause human error.

The primary one is ignoring the workplace policies and/or not being aware of them. Under this case, the employee should presumably not have been aware of the organization's information security policies and the employee should likely have broken the policy resulting in a security incident.

Another reason is that common information security practices would not be known. Awareness of information security includes classification of phishing / spear-phishing emails, malicious email attachments, poor passwords etc. ... Without that basic awareness of information security, the employee becomes a prey to the attacker / hacker. Because currently in cyber field attackers not targeting the company's executive/board members rather they're targeting low-level employees. Recently in the USA, it is reported (Lindsey, 2020) that hackers are targeting the potential employees to spread malware on the organization's system.

The third primary reason is negligence. This could include failing to do the work properly. This factor would mostly be unintentional, but it does have a significant impact on the organization.

**Data breaches occurred due to human error**

Several data breaches have occurred due to human error. The well-known and famous data breach that occurred as a result of human error was Target Data Breach in 2013. Data breach studies (Xiaokui, Danfeng, Ke & Andrew, 2017) indicate that the data breach consists of several human error facts. In the first place, the malware has been planted into the systems in a sophisticated phishing attempt. Second, security warnings have been ignored and no action has been taken against the warnings generated by the monitoring software. According to reports, the total cost of this data breach is nearly 300 Million USD.

## McCumber cube

McCumber cube is a framework that used to manage the protection/Security measures. It basically has 3 sides namely, goals, information state (data at rest, data at transit, data at processing), and security measures (Policies, Education, Training). McCumber cube's security measures model can be applied to the human error as follows,

**Education**

**SETA programs**

SETA stands for Security Education, Education and Awareness. SETA is a training program (Solms, S.H. & Solms, Rossouw. 2009) aimed at educating employees to reduce human error-related data breaches and to increase awareness of information security among them. Implementing SETA programs is one of the recommended (M.G. Lee, 2012) methods and is followed by many organizations at this moment. According to ISO27001 Clause A.8.2 (Calder & Watkins, 2015), The organization should enable this SETA program not only to its employees, but also to its respective contractors and third-party users. Measures to conduct SETA programs at least once a year are recommended.

**Simulated attacks**

Simulated attacks designed by the organizations help in testing the knowledge of safety practices of employees. These are similar to the exercises on fire drill. In simulated attacks, the organization sends the employees phishing / malicious email / email attachments to see how well they are identified and responded to by the employees. This one also widely used in the organization nowadays and one of the recommended methods to follow (Rapid7, 2018).

**Policy**

Policies play a vital role in controlling/preventing the impact of human errors. The most recommended and important policies are password policy and information disclosure policy

**Password policy**

The primary purpose of the password policy is to create a strong password standard, to protect those passwords, and to enable frequently changing passwords. The password policy (Sans, n.d.) applies to all personnel responsible for any account, any system used within the organization, the network, and the facility to store any information. The following is an example of a password policy according to NIST 800-63 (NIST, 2020) guidelines.

- Password should be a minimum of 8 characters when it being set-upped by people
- Password should be a minimum of 6 characters when it being set-upped by a service/system.
- The password should support all the ASCII characters including space.
- Chosen passwords should be checked with a password dictionary.
- Password should support at least 64 characters of maximum length.
- Minimum of 10 attempts before the lockout.

Apart from those NIST framework guidelines, there are some more options which can be added to the password policy,

- Passwords must be changed every 90 days or more frequently.
- The password should contain at least one upper case letter, one lower case letter, one symbol, and one number. This will help to increase the complexity of the password.

**Information disclosure policy**

According to ISO27001 Annex A.7.2 (Humphreys, 2016), organizations should conclude agreements whenever new staff join the organization. It also referred to the Non-Disclosure Agreement, which sets out what information can be disclosed to the public and what information cannot be disclosed. In addition, the agreement should clearly set out the actions to be taken in the event of a breach of the Non-Disclosure Agreement. In addition, it is important that the organization take appropriate measures to make its employees aware of the agreement.

In addition, the organization can use the data classification policy to make this work easier. Data classification helps the organization to classify in terms of confidentiality. A typical data classification has 4 levels (Irwin, 2019) they are, confidential (only higher management must access), Restricted (only particular job roles can access), internal (all the employees can access), public (everyone can access). Classifying the information allows an employee to aware of the information that they can disclose.

Apart from these two policies organizations can consider another policy for BYOD (Bring Your Own Device). Also, organizations are advised to run periodic internal audits/reviews to ensure security measures.

**Technology**

The usage of proper technology will significantly reduce the occurrence of human error. Few technology measures that can be placed to prevent human errors are,

**Employee Monitoring software**

Use of Employee monitoring software is a basic mechanism that can be used by the organization to reduce the occurrence of human error. This facility allows every activity of the employee to be monitored so that, if any security incidents occur, information from the monitoring system can be used to identify the root cause / where it begins. There are several ethical issues associated with this but using this with acceptable policies such as what data can be collected will be beneficial.

**Cryptography and Encryption**

Encryption is the most recommended technology to counteract threats, particularly human error. Organizations should use encryption while resting data (Robb, 2017). Also, the algorithm that is going to be in place should be secure enough, and in general, instead of creating a new algorithm, it's wise to choose one that already exists. Currently, in the industry, there are several vendors like IBM, Dell, McAfee providing their cryptography and encryption products.

**Identity and access management**

Allowing employees to have access only to what they need for their job roles would be an appropriate strategy. Identity and Access Management (IAM) deployment in place would also help to reduce the risk.

**2-factor authentications**

Currently, usage of 2 factor/multi-factor authentication is emerging among the organizations to provide an additional layer of support. According to research reports (Sans & Preston, 2014), the intention of adopting to 2FA is significantly increasing.

## Upcoming landscape

According to the future cyber-attack landscape predictions of the security organization (Checkpoint, 2019), it is estimated that the phishing-based attacks will remain as a top vector of attack and will rise rapidly. Another prediction (Jason, 2019) estimated that the Phishing vector would go beyond e-mail and launch via cloud. So, when the attack vectors evolve the organizations need to keep their tactics up to date to combat them.

## Conclusion

Since this vector of threat is human, it cannot be eliminated, but countermeasures can help to stop human error from turning into potential security breaches that can cause serious impact to the organizations. Technology is also not a single solution in cybersecurity to solve all the problems, but rather the shared responsibility of everyone to keep the digital world safe.

# References

Calder, A., & Watkins, S. (2015). *IT Governance: An International Guide to Data Security and ISO27001/ISO27002*. London: Kogan Page Limited.

Checkpoint. (2019). 2020 Vision: Check Point's cyber-security predictions for the coming year - Check Point Software. Retrieved from https://blog.checkpoint.com/2019/10/24/2020-vision-check-points-cyber-security-predictions-for-the-coming-year/

Humphreys, E. (2016). *Implementing the ISO/IEC 27001* (2nd ed.). Norwood: Artech House.

Irwin, L. (2019). What is information classification and how is it relevant to ISO 27001? - IT Governance UK Blog. Retrieved from https://www.itgovernance.co.uk/blog/what-is-information-classification-and-how-is-it-relevant-to-iso-27001

Jason, C. (2019). Cybersecurity Predictions for 2020. Retrieved from https://www.netskope.com/blog/cybersecurity-predictions-for-2020

Kaspersky. (2017). Retrieved from https://media.kasperskycontenthub.com/wp-content/uploads/sites/100/2017/11/10083900/20170710_Report_Human-Factor-In-ITSec_eng_final.pdf

Lindsey, O. (2020). Iranian Hackers Target U.S. Gov. Vendor With Malware. Retrieved from https://threatpost.com/iran-hackers-us-gov-malware/152452/

M. G. Lee, "Securing the human to protect the system: Human factors in cyber security," 7th IET International Conference on System Safety, incorporating the Cyber Security Conference 2012, Edinburgh, 2012, pp. 1-5.

NIST. (2020). NIST Special Publication 800-63B. Retrieved from https://pages.nist.gov/800-63-3/sp800-63b.html#sec5

Rapid7. (2018). *WHY YOU SHOULD LET YOUR SECURITY TEAM GO PHISHING*. Rapid7. Retrieved from https://www.rapid7.com/globalassets/_pdfs/whitepaperguide/rapid7-whitepaper-why-you-should-let-your-security-team-go-phishing.pdf

Robb, D. (2017). Top 10 Enterprise Encryption Products. Retrieved from https://www.esecurityplanet.com/products/top-encryption-products.html

Sans. Retrieved from https://www.sans.edu/student-files/projects/password-policy-updated.pdf

Sans, & Preston, A. (2014). *Global Information Assurance Certification Paper*. Sans Institute. Retrieved from https://www.giac.org/paper/gsec/35160/impediments-adoption-two-factor-authentication-home-end-users/139464

Solms, S.H. & Solms, Rossouw. (2009). Information Security Education, Training and Awareness. 10.1007/978-0-387-79984-1_10.

Swinhoe, D. (2019). Humans are the weak link: Security awareness & education still a challenge for UK companies. Retrieved from https://www.csoonline.com/article/3430596/humans-are-the-weak-link-security-awareness-education-still-a-challenge-for-uk-companies.html

Verizon. (2019). Verizon: 2019 Data Breach Investigations Report. *Computer Fraud & Security*, *2019*(6). doi: 10.1016/s1361-3723(19)30060-0

Xiaokui, S., Danfeng, Y., Ke, T., & Andrew, C. (2017). Breaking the Target: An Analysis of Target Data Breach and Lessons Learned. *Arxiv*. doi: 701.04940v1